\documentstyle[aps,epsf,amssymb]{revtex}
\begin{document}
\title{Correlated percolation patterns in PEF damaged cellular material}
\author{
N. I. Lebovka $^{a,b}$, 
M. I. Bazhal $^{a,c}$, 
E. Vorobiev $^{a,}$ 
}

\address{
$^{a}$D\'{e}partement de G\'{e}nie Chimique, Universit\'{e} de
Technologie de Compi\`{e}gne, Centre de Recherche de Royallieu,
B.P. 20529-60205 Compi\`{e}gne Cedex, France\\ $^{b}$Institute of
Biocolloidal Chemistry named after F.D. Ovcharenko, NAS of
Ukraine, 42, blvr.Vernadskogo, Kyiv, 252142, Ukraine\\
$^{c}$Ukrainian State University of Food Technologies, 68,
Volodymyrska str., Kyiv, 252033, Ukraine\\ }
\maketitle
\begin{abstract}
\tighten We present results of numerical and experimental
investigation of the electric breakage of a cellular material in
pulsed electric fields (PEF). The numerical model simulates the
conductive properties  of a cellular material by a two-dimensional
array of biological cells. The application of an external 
in the form of the idealised square pulse sequence with a pulse
duration $t_{i}$, and a pulse repetition time $\Delta t$ is
assumed. The simulation model includes the known mechanisms of
temporal and spatial evolution of the conductive properties of
different microstructural elements in a tissue. The kinetics of
breakage at different values of electric field strength $E$,
$t_{i}$ and $\Delta t$ was studied in experimental investigation.
A 5 mm x 55 mm cylindrical slab of apple is taken as a sample. The
results of the experimental and numerical studies were compared.
We propose the hypothesis for the nature of tissue properties
evolution after PEF treatment and consider this phenomena as a
correlated percolation, which is governed by two key processes:
resealing of cells and moisture transfer processes inside the
cellular structure. The breakage kinetics was shown to be very
sensitive to the repetition times $\Delta t$ of the PEF treatment.
We observed correlated percolation patterns in a case when $\Delta
t$ exceeds the characteristic time of the processes of moisture
transfer and random percolation patterns in other cases. The
long-term mode of the pulse repetition times in  PEF treatment
allows us to visualize experimentally the macroscopic percolation
channels in the sample. We observe considerable differences
between the damage kinetics at long and short repetition times
both for experimental and simulation data.

\end{abstract}

\bigskip\bigskip
\tighten {Keywords: Pulsed electric fields; Computer simulation;
Electroporation; Resealing; Moisture transfer; Percolation;
Apples}

\tableofcontents

\bigskip\bigskip


\begin{tabular}{ll}
{\bf Notation} 

\\$A$ & $\sim 4d_{c}^{2}$, cross section area of a single cell or a
membrane, m$ ^{2}$

\\ $c$ & $=r_{c}^{f}/r_{c}^{i}=\sigma_{c}^{i}/\sigma_{c}^{f}\leq 1$,
resistance moisture transfer coefficient

\\ C$_{m}$ & specific capacity of membrane, F m$^{-2}$

\\ C$^{\ast }$ & = C$_{m}$($\varepsilon_{w}$/$\varepsilon _{m}$-1)/(2$\gamma $)

\\$d_{m}$ & membrane thickness, m

\\ $2d_{c}$ & cell diameter, m

\\$E$ & electric field strength, kV cm$^{-1}$

\\$E^{\ast }$ & $=2d_{c}E/u_{o}$, normalized electric field strength

\\ $\Delta F^{\ast }$ & $=\pi \omega ^{2}/(kT\gamma )$, reduced critical free
energy of pore formation

\\ $G$ & conductivity of the bond in the network, Ohm$^{-1}$

\\ $k$ & Boltzmann constant, $1.381\times10^{-23}$ J K$^{-1}$

\\ $L$ & $=2d_{c}N$ , total thickness of a sample (slab of apple), m

\\ $m$ & $=r_{m}^{i}/r_{m}^{f}=\sigma_{m}^{f}/\sigma_{m}^{i}$,
membrane resistance resealing coefficient

\\ $n$ & number of pulses

\\ $N$$\times$$N$ & dimensions of a 2D lattice

\\ $P$ & degree of biological tissue damage

\\ $r$ &  resistance of the model resistor in the network, Ohm

\\ $r_{m}$ &  membrane part of $r$ resistance, Ohm

\\ $r_{c}$ &  cellular part of $r$ resistance, Ohm

\\$S(u)$ &  survival probability function


\\$t_{i}$ & pulse duration, $\mu $s

\\ $dt$ & impact time duration, or "elementary" time step $dt\sim
0.1 t_{i}$, s

\\ $\Delta t$ & pulse repetition time, ms

\\ $T$ & temperature, K

\\ $u$ & transmembrane voltage, V

\\ $u_{o}$ & midpoint of a survival probability function $S(u)$, V

\\ $u^{\ast }$ & $=u/u_{o}$, normalized transmembrane voltage

\\ $\Delta u$ & width of a survival probability function $S(u)$, V

\\ $\Delta u^{\ast }$ & $=\Delta u/u_{o}$, normalized width of a
survival probability function $S(u)$

\\ $U$ & external voltage, V

\\ $W$ & moisture content, \%

\end{tabular}


\begin{tabular}{ll}

\bigskip \\


{\it Greek letters}

\\$\varepsilon _{w}$ & $=80$, dielectric constant of water

\\$\varepsilon _{m}$ & $=2$, dielectric constant of membrane

\\$\gamma $ & surface tension of membrane, N m$^{-1}$

\\$\lambda $ & adjustable relaxation parameter

\\ $\sigma $ & conductivity, S m$^{-1}$

\\ $\sigma _{c}$ & conductivity of cellular material (barring membrane), S m$^{-1}$

\\ $\sigma _{m}$ & conductivity of membrane, S m$^{-1}$

\\ $\tau_{m} $ & lifetime of a membrane, s

\\ $\tau_{\infty }$ & parameter, lifetime of a membrane at $T=\infty$, s

\\ $\tau_{r} $ & characteristic time of membrane resealing, s

\\ $\tau_{d} $ & characteristic time of a moisture
transfer processes after PEF treament, s

\\ $\omega $ & linear tension of membrane, N

\end{tabular}

\begin{tabular}{ll}

\bigskip
\end{tabular}

{\it Superscripts}

\begin{tabular}{ll}
$d$ & damaged

\\ $i$ & initial

\\ $f$ & final

\bigskip
\end{tabular}

{\it Subscripts}

\begin{tabular}{ll}
$c$ & cellular

\\$e$ & effective

\\ $i$ & intact cell

\\ $j$ & juice


\\ $m$ & membrane

\\ $t$ & total
\bigskip
\end{tabular}

{\it Abbreviations}

\begin{tabular}{ll}
CEF & continuous electric field

\\ PEF & pulsed electric field
\end{tabular}


\section{Introduction}

Among different nonthermal processing methods used in food
technologies, the pulsed electric field (PEF) treatment is one of
the most promising. A number of new PEF applications were
demonstrated for anti-microbial treatment of liquid
foods, e.g., fruit juices, milk etc., 
(Barbosa-C\'{a}novas, Pothakamury, Palou \& Swanson, 1998;
Barsotti \& Cheftel, 1998; Wouters \&
Smelt, 1997), 
and for the cellular tissue materials 
(Knorr \& Angersbach, 1998, Knorr, Geulen, Grahl \& Sitzmann,
1994). For years back, the continuous electric field (CEF)
treatment was also shown to be good for juice yield
intensification and
for increasing the product quality in juice production 
(Bazhal \& Vorobiev, 2000; McLellan, Kime \& Lind, 1991; Scheglov,
Koval, Fuser, Zargarian, Srimbov, Belik et al., 1988),
processing of vegetable and plant raw materials 
(Papchenko, Bologa \& Berzoi, 1988; Grishko, Kozin \& Chebanu,
1991), processing of foodstuffs (Miyahara, 1985),
winemaking (Kalmykova, 1993), and sugar production 
(Gulyi, Lebovka, Mank, Kupchik, Bazhal, Matvienko et al., 1994;
Jemai, 1997). But all these CEF applications were restricted by
high and uncontrolled increases in food temperature.

Extension of different PEF applications to nonthermal processing
of heterogeneous food materials is limited today by the absence of
criteria for choosing optimal parameters of PEF treatment and the
unclear mechanism of electric breakdown processes in the cellular
systems. Recently, a significant advance in 
understanding of the nature and mechanisms of electric field
influence on different animal, plant, and microbial cells has
occurred (Chang, Chassy, Saunders \& Sowers, 1992; Weaver \&
Chizmadzhev, 1996). The strong electric field causes
electroporation of cells, increase of their permeability, and, in
some cases, disruption of their structural integrity (Zimmermann,
1975). The PEF parameters (field strength $E$,pulse duration
$t_{i}$ and number of pulses $n$) can influence both the degree of
membrane destruction or structural alteration and the density of
pores in membrane (Rols \& Teissie, 1998; Gabriel \& Teissie,
1999). The electroporation became very popular because it was
found to be an exceptionally practical way of transferring drugs,
genetic material (e.g. DNA), or other molecules inside the
cells (Chang, Chassy, Saunders \& Sowers, 1992;  
Neumann, Kakorin \& T\oe nsing, 1999). This phenomena is sometimes
 also called as electropermeabilization.

For complex material, such as  tissue, cellular or food material,
PEF application results in increase in the electric conductivity
and permeability of the whole sample. But the nature of
electropermeabilization of complex cellular materials is not yet
well understood in all details. The cellular materials are highly
heterogeneous and the electrical properties of such systems depend
on the electrical properties of single cells as well as on the
geometrical and topological properties of materials (Sahimi,
1994). Here the percolation phenomena may play an important role
for interpretation of the observed experimental results.
Particularly, there is no simple relation between the degree of
material damage and its electrical conductivity.

A number of ambiguous and yet unexplained phenomena are observed
in this field. For example, long-term changes in the conductivity
of a cellular material after its electric field treatment are
usually observed. It was reported, for example, that the
conductivity of the vegetable tissue can decrease after
termination of electric treatment over at least 24 hours
(Kulshrestha \& Sastry, 1998). The explanation of this phenomenon
is not trivial since multiple mechanism can be responsible for
time evolution of the conductivity.

One of possible explanations is based on the assumption about
existence of a partial membrane resealing after its breakage in a
high electric field. Note, that the electric damage of a cell is
itself a rather complex process and there may exist different time
scales for the kinetics of pore evolution. At low strength
electric field, $E$ ($<200$ V cm$^{-1}$), and short pulse
duration, $t_i$ ($\sim 10^{-5}-10^{-6}$ s) the electrical
breakdown is spontaneously reversible, and all the damages
disappear after the field is switched out (Abidor, Arakelyan,
Chernomordik, Chizmadzhev, Pastushenko \& Tarasevich, 1979; Weaver
\& Chizmadzhev, 1996). At moderate PEF treatment ($E=0.5-2$ kV
cm$^{-1}$, $t_i\sim 10^{-4}-10^{-5}$ s) the integrity of cells
drops rapidly, but due to resealing or recovering process some of
the cells loose their permeability and the pores may persist in
the membrane at larger after PEF application. The resealing
process time constant, $\tau_r$ may be very large, of order
$1-10^2$ s at 25 $^{\circ}$C (Neumann \& Boldt, 1990; Chang \&
Reese, 1990;
Chizmadzhev, Indenbom, Kuzmin, Galichenko, Weaver \& Potts,
1998; 
Neumann, T\oe nsing, Kakorin, Budde \& Frey, 1998; Weawer,
Pliquett \& Vaughan, 1999; DeBruin \& Krassowska, 1999). For
vegetable food materials the reported resealing time constant was
of order 1 s (Knorr, Heinz, Angersbach \& Lee, 2000). The high PEF
treatment ($E=10-50$ kV cm$^{-1}$, $t_i\sim 10^{-6}$ s) causes the
irreversible damages and this mode of treatment is used for
inactivation of microorganisms (Barbosa-C\'{a}novas et al., 1998).
The time constant of a resealing process is a complex function of
$E$ and $t_i$ values and depends on the type of cells or
membranes. A number of mechanisms were proposed  for explanation
of resealing processes, but in a general case the nature of the
long-lived permeabilization is still unclear (Saulis, 1997;
Teissie \& Ramos, 1998; Chizmadzhev et al., 1998; Weaver et al.,
1999).

Another cause of the long-term changes in the conductivity may be
related with the different transport phenomena in a structured
cellular material (Aguilera \& Stanley, 1999),  e.g. diffusional
motion, osmotic flow and redistribution of moisture inside the
sample, which can be enhanced by PEF application. We can estimate
the time constant of the diffusion processes inside the cellular
material as $\tau_{d}\sim d_{c}^{2}/(6D)\approx 1$ s at
25$^{\circ}$ C, where $d_{c}\sim 10^{-4}$ m is a radius of cell,
and $D\sim 10^{-9}$ m$^{2}$ s$^{-1}$ (Gekkas, 1992) is an
effective diffusion coefficient for the moisture inside a cellular
material.


The aim of this study is to 
elucidate the mechanism of PEF treatment and long-term changes in
the conductivity of cellular materials. We consider the damage of
a biological tissue in the electric field as a correlated
percolation phenomena, which is governed by the resealing and
moisture transfer processes. The developed simulation model 
includes the known mechanisms of temporal and spatial evolution of
the conductive properties of microstructural elements (cell
membranes, tissue frameworks, etc.). The results of the
experimental and numerical studies are compared and possible
scenarios of the tissue conductive properties evolution after PEF
treatment are discussed.

\section{Materials and experimental methods}

\subsection{Materials}
\label{Materials}

 Freshly harvested apples of the Golden Delicious
variety were selected for investigation and stored at $4{{}^{\circ
}}$C until required.
A moisture content of apples 
$W$ was within 80-85\%. Typical high resolution scanning electron
micrograph of the apple sample is presented in Fig. \ref{f0}.
Images were obtained on the instrument XL30 ESEM-FEG (Philips,
V=15 kV, P=3.5 Torr). The initial specific conductivity of samples
(before treatment) was within the interval  $\sigma^{i}=0.003 -
0.007$ S m$^{-1}$. The final specific conductivities (after
treatment) depend upon the mode of treatment and they were within
the range of $\sigma^{f}=0.035 - 0.070$ S m$^{-1}$. The specific
conductivity of the apple juice extracted from the sample apples
was $\sigma_{j}=0.22 \pm 0.05$ S m$^{-1}$.

\subsection{Experimental methods}

The conductivities were measured by contacting electrode method
with an LCR Meter HP $4284$A (Hewlett Packard, 38 mm guarded/guard
Electrode-A HP 16451B) for thin apple slice samples at a frequency
of $100$ Hz and with a Conductimetre HI$8820$N (Hanna Instruments,
Portugal) for the apple juice samples at a frequency of $1000$ Hz
(these frequencies were selected as optimal in order to remove the
influence of the polarizing effects on electrodes and inside the
samples).

Figure \ref{f1} is a schematic representation of the experimental
pulsed electric field treatment set-up. A high voltage pulse
generator, $1500$V-$15$A (Service Electronique UTC, France)
allowed to vary $t_{i}$  within the interval of $10-1000$ $\mu $s
(with precision $\pm 2$ $\mu $s), $\Delta t$ within the interval
of $1-100$ ms (with precision $\pm 0.1$ ms) and $n$ within the
interval of $1-100000$.

Pulse protocols and all the output data (current, voltage,
impedance and temperature) were controlled using a data logger and
a special software HPVEE v.4.01 (Hewlett-Packard) adapted by
Service Electronique UTC, France. The temperature was recorded in
the on-line mode by a thermocouple THERMOCOAX type 2 (AB 25 NN,
$\pm $0.1${^{\circ }}$C).

The thin apple slabs (of thickness $5\pm 0.2$ mm  and of diameter
$55\pm 0.5$ mm) were used as samples in the present investigation.
The freshly cut cells at the outer boundary of a sample cause the
initial time dependence of a sample conductivity. During the
period about 200 s (Fig. \ref{f2}) the conductivity achieves about
90 \% of its stationary value. So, in all our experiments we
always skipped this transition time before the PEF treatment.
Usually the sample conductivity was measured 10 s after finishing
of treatment procedure. All experiments were repeated at least
five times.

\section{Description of the simulation model}

\subsection{Probability of a single cell damage}

Weaver \& Chismadzhev (1996) gave the comprehensive discussion of
different models of the membrane rupture. The model that seems to
be most reasonable from the physical point of view, is the, so
called, transient aqueous pore model, in which the average
membrane lifetime $\tau_{m}$
can be estimated with the help of the following expression:

\begin{equation}
\tau_{m}(u) =\tau _{\infty }\exp (\Delta F^{\ast }/(1+u^{2}C^{\ast
})). \label{e01}
\end{equation}
where $\tau _{\infty }$ is a parameter (equals to a lifetime of a
membrane at infinite temperature, $T=\infty$), $\Delta F^{\ast
}=\pi \omega ^{2}/(kT\gamma )$ is a reduced critical free energy
of pore formation, $k=1.381\times10^{-23}$ J K$^{-1}$ is the
Boltzmann constant, $\omega$ is a linear tension of membrane, N,
 $\gamma $ is a surface tension of membrane, N m$^{-1}$, $u$ is a
transmembrane voltage, V, $C^{\ast} = C_{m}(\varepsilon
_{w}/\varepsilon _{m}-1)/(2\gamma) $, $C_{m}$ is a specific
capacity of membrane, F m$^{-2}$, $\varepsilon _{w} =80$ is a
dielectric constant of water, $\varepsilon _{m}=2$ is a dielectric
constant of membrane.

Then the survival probability for a membrane (as a whole cell)
during the impact period of $dt$ may be estimated as

\begin{equation}
S(u)=\exp (-dt/\tau_{m}(u)).  \label{e02}
\end{equation}

Taking Eq. (\ref{e01}) into account, we can rewrite Eq.
(\ref{e02}) in the following convenient dimensionless form

\begin{equation}
S(u^{\ast })=\exp \left( -\ln 2/\exp a([1-(1-u^{\ast }{}^{2})/
(a\Delta u ^{\ast }\ln 2)]^{-1}-1)\right) ,  \label{e03}
\end{equation}
where $u^{\ast }=u/u_{o}$, $\Delta u^{\ast }=\Delta u/u_{o}$,
$u_{o}=\sqrt{(\Delta F^{\ast }/a-1)/C^{\ast }}$, $\Delta
u=u_{o}/\left( (1-a/\Delta F^{\ast })a\ln 2\right)$, and $a=\ln
(dt/(\tau_{\infty }\ln 2))$ is a parameter.

There is no first principle basis for correct estimations of the
different parameters in Eqs. (\ref{e01})-(\ref{e03}) (Weaver \&
Chismadzhev, 1996), so the numerical values obtained from fitting
of $\tau_{m}(u)$ to experimental data are used  as a rule. For
example, Lebedeva (1987) presented the following estimations for
the lipid membranes: $\tau _{\infty }\approxeq 3.7\times10^{-7}$ s
, $\omega \approxeq
1.69\times10^{-11}$ N, $\gamma \approxeq 2\times10^{-3}$ N m$^{-2}$, $%
C_{m}\approxeq 3.5\times10^{-3}$ F m$^{-2}$ at 25 $^{\circ}$C.
>From these estimations  we can obtain the following parameters for
Eq. (\ref{e03}): $u_{o}\approxeq 1.52$ V, $\Delta u^{\ast
}\approxeq 1.07$ and $a\approxeq 1.36 $ (at $dt=1$ $\mu $s) and
$u_{o}\approxeq 0.71$ V, $\Delta u^{\ast }\approxeq 0.26$ V, and
$a\approxeq 5.97$ (at $dt=100$ $\mu $s). But for real cellular
systems the parameters of Eq. (\ref{e03}) are not clear-cut and
they would depend on physical properties, type and quality of raw
materials as well as on the value of $dt$. Yet, for definiteness
in the following computation we always use the parameters of $\tau
_{\infty }$, $u_{o}$ and $\Delta u^{\ast }$ estimated on the basis
of aforecited data of Lebedeva (1987).

The example of the survival curve $S(u^{\ast })$ is shown in Fig.
\ref{f3}. We see that $S(u^{\ast })$ is a kind of probability
transition function and $u^{\ast }=1$ ($u=u_{o}$) corresponds to
the midpoint, where $S(u)=1/2$. In fact, the value of $u_{o}$ may
serve as an estimate for the critical value of transmembrane
voltage, which causes the abrupt decrease of the membrane
lifetime. Here we define the width of this transition function
$\Delta u^{\ast }$ by drawing a tangent straight line to a curve
$S(u^{\ast })$ in the midpoint $u^{\ast }=1$á as it is shown in
the Fig. \ref{f3} (see dotted line). The dashed line at this
Figure corresponds to the normalised density distribution function
$S^{\prime}/S^{\prime}{_{max}}$, where $S^{\prime}=dS/du^{\ast}$.

\subsection{Simulation procedure}

\subsubsection{Resistor network model}

We simulate the conductive structure of a cellular material as a
two-dimensional array of cells located at the nodes of a simple
square lattice. The lattice has a size $N^2$ with periodic
boundary conditions in the $x$ direction in order to reduce the
finite size scaling effects (Watanabe, 1995). The boundary
conditions for $y$ direction are as follows: at $y=0$ and $y=N+1$
we put two electrodes with a constant potential difference $U$
(Fig. \ref{f4}). So the mean drop of potential per cell is equal
to $u=U/(N+1)$ and the reduced voltage on membrane in Eq.
(\ref{e03}) is defined as $u^{\ast}=U/(u_{o}(N+1))$. Then mean
strength of the electric field along $y$-axis is equal to
\begin{equation} E=U/(2d_{c}(N+1))=u^{\ast}u_{o}/(2d_{c}),
\label{e04}
\end{equation}
where $2d_{c}$ is a cell diameter.

>From this equation we obtain, for example,
$E=u^{\ast}u_{o}/2d_{c}= 50u^{\ast}$ V cm$^{-1}$ at $u_{o}=1$ V
and $d_{c}= 100$ $\mu$m. We can introduce also the normalised
field strength defined as $E^{\ast }=2d_{c}E/u_{o}\equiv
u^{\ast}$. As it was mentioned above, the exact value of $u_{o}$
unknown. So we can chose the $u_{o}$ parameter from the condition
of best fitting to the observed experimental data. In our
simulation we use $N=250$ and the total sample thickness is
$L=2d_{c}N$ ($\approx 5$ cm when $d_{c}\approx 100$ $\mu$m).

\subsubsection{Microstructural conductive properties}
\label{Microstructural}

We suppose that each node is connected with neigbouring nodes
through four conducting resistors, which simulate the conductive
properties of the cellular media microstructural elements. The
resistance of such resistors is determined by the two constituent
parts
\begin{equation}
r=r_{m}+ r_{c},  \label{e05}
\end{equation}
which correspond to the membrane ($r_{m}$) and cellular ($r_{c}$)
medium contributions, respectively. Here, membrane contribution
includes the effective conductive properties of  the different
membranes in the cellular structure (mainly plasmatic and
tonoplast membrane). Cellular medium contribution reflects both
intra- and extra-cellular conductive properties of cellular
materials. Intra-cellular contribution includes the effective
conductive properties of  the cytoplasm with its organelles
(occupies about 10 \% of the cell volume), and the vacuole (about
80 \% of the cell volume). Extra-cellular contribution includes
the apparent conductive properties of the rigid cell wall
(occupies about 10 \% of the total volume and its main structural
element is cellulose), of pores and intercellular spaces filled
with air etc., (account for around 20-25 \% of the total volume in
apple, see Aguilera  \& Stanley (1999)).

We estimate the resistance values of $r_{m}$ and $r_{c}$ in Eq.
(\ref{e05}) as
\begin{equation}
r_{m}=d_{m}/(\sigma_{m} A),  \label{e06}
\end{equation}
and
\begin{equation}
r_{c}=d_{c}/(\sigma_{c} A),  \label{e07}
\end{equation}
where $d_{m}$ is the thickness of membrane, $A$ is a mean
cross-section area of a single cell, $\sigma_{m}$  and
$\sigma_{c}$ are the conductivities of the membranes and cellular
material (barring membranes), respectively.

At the initial stage of simulation we suppose that all the cells
are intact and corresponding resistors in the model are equal to
\begin{equation}
r^{i}=r_{m}^{i}+r_{c}^{i}.  \label{e08}
\end{equation}

In this case the effective conductivity of the whole sample
$\sigma$ may be calculated as
\begin{equation}
\sigma=\sigma^{i}=\frac{d_{m}+d_{c}}{r
A}=\frac{d_{m}+d_{c}}{d_{m}/\sigma_{m}^{i}+d_{c}
/\sigma_{c}^{i}}\simeq\frac{\sigma_{m}^{i}d_{c}/d_{m}}{1+\sigma_{m}^{i}
d_{c}/\sigma_{c}^{i}d_{m}},  \label{e09}
\end{equation}
where we take into account that $d_{m}\ll d_{c}$.

If the potentials in all the nodes are known, $u_{x,y}$, then we
can easily determine the transmembrane voltages $u$ at all
membranes in a system. Consequently, we can determine with the
help of Eq. (\ref{e03}) which of membranes will destroy after the
PEF treatment. The conductivity of these membranes after PEF
breakage increases considerably ($\sigma_{m}^{i}\Rightarrow
\sigma_{m}^{d}\rightarrow \infty$), and so $r_{m}^{i} \Rightarrow
r_{m}^{d}\simeq 0$).

Figure \ref{f5}(a) presents the case, when the potential
difference in the vertical ($y$) direction $|u_{x,y}-u_{x,y+1}|$
exceeds some critical value and, as a result, two cells in the
vertical direction are damaged. Note, that at all accounts we
always observe for this model the simultaneous damage of two
cells, because two adjacent cells suffer equal voltage loading. In
this case, we should make the following interchange of the
resistors
\begin{equation}
r^{i}\Rightarrow r^{d}=r(t)= r_{m}(t)+r_{c}(t),
 \label{e14}
\end{equation}
as it is shown at the Fig. \ref{f5}(a). The similar case for
horizontal ($x$) direction is presented in Fig. \ref{f5}(b). Here
$r^d$ corresponds to resistance of damaged cell, and its time
evolution may be found with the help of Eqs.
(\ref{e10})-(\ref{e14}).

\subsubsection{Resealing processes}
\label{Resealing}

The model accounts for the possibility of temporal
electropermeabilization as follows. If any membrane is damaged
then it begins to reseal immediately, and we suppose, that this
resealing results in increasing of the $r_{m}$ as
\begin{equation}
r_{m}(t)\simeq r_{m}^{f}(1-e^{-t/\tau_{r}}) \label{e10}
\end{equation}
where $r_{m}^{f}=d_{m}/(\sigma_{m}^{f}A$), $\sigma_{m}^{f}$ is a
final conductivity of a membrane after the complete resealing, and
$\tau_{r}$ is a time constant of resealing process.

We define the  membrane resealing coefficient  as
\begin{equation}
m=r_{m}^{i}/r_{m}^{f}=\sigma_{m}^{f}/\sigma_{m}^{i}\leq 1
\label{e11}
\end{equation}

\subsubsection{Moisture transfer processes}
\label{Moisture}

The moisture transfer processes at different hierarchical levels,
such as diffusional migration, osmotic flow and redistribution of
moisture inside the sample (Aguilera \& Stanley, 1999), enhance as
a result of PEF application. The new conducting channels arise
inside the sample and this causes the temporal decreasing of
$r_{c}$ value. We approximate this evolution as
\begin{equation}
r_{c}(t)=r_{c}^{i}-(r_{c}^{i}-r_{c}^{f})(1-e^{-t/\tau_{d}})
\label{e12}
\end{equation}
where $r_{c}^{f}=d_{c}/(\sigma_{c}^{f} A)$, $\sigma_{c}^{f}$ is a
final conductivity of a cellular material after completion of
moisture transfer process in the sample, and $\tau_{d}$ is a time
constant of this process.

For the quantitative description of the moisture transfer
processes contribution to the change in cellular material
conductivity we introduce the resistance moisture transfer
coefficient $c$ defined as
\begin{equation}
c=r_{c}^{f}/r_{c}^{i}=\sigma_{c}^{i}/\sigma_{c}^{f}\leq 1
\label{e13}
\end{equation}

\subsubsection{Finite element analysis}
\label{Finite}

 The simulation of temporal evolution of
the system
requires a knowledge of the potential distribution in the lattice.
This distribution can be obtained numerically by solving (Lebovka
 \& Mank, 1992) the discretized version of Laplace's equation
on a lattice with given boundary potentials. For this purpose we
have used the 
successive relaxation scheme (Press et al.,
1997).
We update the chosen site potential $u_{x,y}^{n}$ at the n-th
relaxation step according to the following equation
\begin{equation}
u_{x,y}^{n}=u_{x,y}^{n-1}+\lambda \left( \frac{
G_{1}u_{x,y+1}^{n-1}+G_{2}u_{x-1,y}^{n-1}+G_{3}u_{x,y-1}^{n-1}+G_{4}u_{x+1,y}^{n-1}
}{G_{1}+G_{2}+G_{3}+G_{4}}-u_{x,y}^{n-1}\right) \label{e15}
\end{equation}
where $\lambda$ is an adjustable relaxation parameter and
$G_{1}\div G_{4}$ are the conductivities of the bonds which
connect the chosen site $x,y$ with all its neighbours (Fig.
\ref{f6}).

The iteration procedure over all sites in the lattice is continued
until the maximum of the relative difference between potentials in
two successive iterations, $u_{x,y}^{n}/u_{x,y}^{n-1}-1$,
converges to a small value $\delta$ ($=10^{-3}$).

\subsubsection{Simulated properties and main parameters}
\label{Simulated}

The effective media conductivity 
$\sigma$ was calculated on the basis of $r(x,y)$ values by
applying 
a highly efficient Frank \& Lobb
(1988) algorithm. The total damage degree $P$ was estimated as the
membrane damage degree with the help of the following relation
\begin{equation}
P=\left( 1-\frac{1}{4N^{2}}\sum\limits_{x,y=1}^{N}r_{m}(x,y)
/r_{m}^{i}\right) \label{e16}
\end{equation}

Note that $P=1$ when all cells are damaged ($r_{m}(x,y)\equiv
r_{m}^{d}= 0$) and $P=0$ when all cells are intact
($r_{m}(x,y)\equiv r_{m}^{i}$).

We assume the pulse application of external electric field in the
form of an idealised square pulse sequence with a pulse duration
$t_{i}$, and a pulse repetition time $\Delta t$. In order to
increase the accuracy of calculation we introduce the "elementary"
time step $dt$ which is much smaller then the pulse duration
$t_{i}$. In this work we put $dt=0.1t_{i}$.

We use in our calculations the following values of parameters: $
d_{m}=5\times10^{-9}$ m, $d_{c}=10^{-4}$ m,
$\sigma_{m}^{i}=3\times10^{-7}$ S m$^{-1}$ (Kotnik, Miklavcic \&
Slivnik, 1998), $\sigma _{c}^{f}=0.1$ S m$^{-1}$ (approximately
corresponds to the conductivity of the absolutely damaged cellular
material), and treat $m$, $c$, $\tau_{r}$ and $\tau_{d}$ as
adjustable variables.

With this sets of parameters we can adjust the experimentally
observed parameters $\sigma^{i}$  and $\sigma^{f}$ (see Section
(\ref{Materials})). For example, in the case when $c=0.1$, we
obtain $\sigma^{i}\simeq 3.75\times10^{-3}$ S m$^{-1}$ (Eq.
\ref{e09}), and $\sigma^{f}\simeq 0.100$ S m$^{-1}$ at $m=0$
(i.e., when resealing is absent) and $\sigma^{f}\simeq
1.07\times10^{-2}$ S m$^{-1}$ at $m=0.5$.

The example of the simulated kinetics of a breakdown is presented
in Fig. \ref{f7}. During the period of pulse action we observe
destruction of a system and increase of $P$ and $\sigma$.  In the
interpulse period the system begins to reseal and we can observe
the partial decrease of $P$ and $\sigma$.

\section{Results and discussion}
\label{Results}

\subsection{Experimental results}
\label{Experimental}

\subsubsection{Damage kinetics}
\label{ExperDamageKinetics}

Figure \ref{f8} presents the examples of the experimental curves
of apple slabs relative conductivity
$\protect\sigma^{f}/\protect\sigma ^{i}$ versus time $t$
dependencies at different values of the electric field strength
$E$ and pulse protocols: $t_i=1$ ms, $n=1-15$, $\Delta t=60$ s.
After each pulse application we observe a rather long-time
resealing-like behaviour of $\sigma^{f}/\sigma^{i}$ values during
the period of order 10 s. So, in each case we measured the
equilibrium values of $\sigma^{f}/\sigma^{i}$ at time
$t_{m}\approx 10$ s after each PEF pulse. The results of
measurement at two pulse protocols $t_i=1$ ms, $n=1-15$, $\Delta
t=60$ s (protocol I), and $\Delta t=10$ ms (protocol II) and
different values of the electric field strength $E=200$ V
cm$^{-1}$ and $E=500$ V cm$^{-1}$ are presented in Fig. \ref{f9}.
We see, that there exist significant difference between kinetics
of $\sigma^{f}/\sigma^{i}$ for these two pulse protocols. As we
have demonstrated before (Lebovka et al, 2000) the pulse
repetition time in the interval of  $\Delta t=1-100$ ms does not
influence the $\sigma^{f}/\sigma^{i}$ versus $n$ dependencies
essentially. 
We have enlarged significantly the pulse repetition time in the
protocol I ($\Delta t=60$ s) 
which resulted in the significant elevation of the conductivity
curves 
to compare with those obtained for the protocol II.

\subsubsection{Visualization of damages}
Figure \ref{f10} shows the photographs that illustrate the
macroscopic structure changes of the apple slabs after PEF
treatment using protocols I(a) and II(b). The dark (brown) spots
on the slabs treated using the protocol I seems to correspond the 
formation of the moisture-saturated and more conductive channels
in the cellular material, which are absent in the case of the
protocol II . The visually observed behaviour reveals the
different modes of cellular material breakage. The only difference
between protocols I and II is the pulse repetition time $\Delta
t$. The considerable increasing of $\Delta t$ in the case of
protocol II allows us to visualize the existence of certain
long-time and large scale moisture transfer processes.

\subsection{Numerical results}
\label{Numerical}

\subsubsection{Simulation of damage kinetics}
\label{SimulDamageKinetics}

Figure \ref{f11}(a) represents some examples of the simulated
$\sigma^{f}/\sigma^{i}$ kinetic curves after application of $n$
($n=1-15$) pulses of $t_i=1$ ms duration for two different pulse
repetition times $\Delta t=60$ s (solid lines, protocol I) and
$\Delta t=10$ ms (dashed lines, protocol II). We have used in
these calculations the following values of parameters: $m=0.1$,
$c=0.1$, $\tau_{r}=10$ s, $\tau_{d}=0.5$ s and $E^{*}=0.65$. For
protocol I we have $\Delta t > \tau_{r}, \tau_{d}$, which means that 
the resealing and mass transfer processes are finished 
during the 
time interval $\Delta t$ and we observe the typical transition
process. For protocol II we have $\Delta t\ll \tau_{r}, \tau_{d}$,
and in this case the system is in transient state during the PEF
treatment. For this case we observe the increase of the relative
conductivity during the time period of order $\tau_{d}$ resulting
from moisture transfer processes and subsequent decrease of this
value due to resealing processes.

Figure \ref{f11}(b) depicts the $(\sigma^{f}/\sigma^{i})_{t_{m}}$
versus $n$ dependencies for the 
above-mentioned  protocols. The calculated values of
$(\sigma^{f}/\sigma^{i})_{t_{m}}$ were "measured" at time $t_{m}$
after $n$-th PEF pulse application, and this procedure corresponds
to the real experimental measurement procedure described earlier
in Section \ref{ExperDamageKinetics}. So, we can consider the Fig.
 \ref{f11}(b) as a simulated analogue of Fig. \ref{f9}. The
kinetics of $(\sigma^{f}/\sigma^{i})$ for the protocol I is
represented in Fig. \ref{f11}(b) by a dashed-dotted line. It is
evident from these data that increase of the pulse repetition time
$\Delta t$ leads to the elevation of
$(\sigma^{f}/\sigma^{i})_{t_{m}}$ versus $n$ dependencies in
accordance with the experimental observations as discussed in
Section \ref{ExperDamageKinetics}.

The examples of simulated breakage patterns for two different
pulse repetition times $\Delta t=60$ s (protocol I) and $\Delta
t=10$ ms (protocol II) are shown in Figs. \ref{f12}(a) and (b)
respectively. Here, each pattern displays only those cells, which
were broken after the $n$-th pulse. We can see that the long
repetition time pulse protocol I results in more extended and
spatially correlated damage patterns (Fig. \ref{f12}(a)). Dark
clusters of the damaged cells show clearly the existence of
collective percolation phenomena, which are typical for the
electrical breakage of inhomogeneous materials (Sahimi, 1994). The
short repetition time pulse protocol II results in more rare and
uncorrelated damage patterns (Fig. \ref{f12}(b)).

\subsubsection{Effects of resealing and moisture transfer processes}
\label{SimulResMass}

The influence of $m$ and $c$ parameters on the "measured" values
of $(\sigma^{f}/\sigma^{i})_{t_{m}\rightarrow\infty}$ after the
application of $n=15$ pulses ($\tau_{d}=0.5$ s, $E^{*}=0.65$) are
demonstrated in Figs. \ref{f13}(a) and (b), respectively. Here,
Fig. \ref{f13}(a) presents
$(\sigma^{f}/\sigma^{i})_{t_{m\rightarrow\infty}} $ versus $\Delta
t$ dependencies for the case of $c=0.1$ and $\tau_{r}=10$ s, and
Fig. \ref{f13}(b) presents the analogue dependencies for the case
of $\tau_{r}=\infty$ (i.e., when resealing is absent). All these
dependencies display characteristic dispersion behaviour in the
range where $\Delta t\sim \tau_{r}$, and in all the cases we
observe increase of $(\sigma^{f}/\sigma^{i})_{t_{m}}$ with $\Delta
t$ increase.

The results for the simulated kinetics of damage degree $P$ and
relative conductivity $\sigma^{f}/\sigma^{i}$ for different values
of $\tau_{d}$ are presented in Fig. \ref{f14}(a) ($\tau_{r}=10$ s)
and Fig. \ref{f14}(b) ($\tau_{r}=\infty$). These calculations were
performed for the following values of parameters: $m=0.05$,
$c=0.1$ s and $E^{*}=0.65$. The data show that resealing
influences significantly the $\sigma^{f}/\sigma^{i}$ versus $t$
dependencies and this effect is the mostly pronounced at large
values of $\tau_{d}$. At small $\tau_{d}$ values ($< 5$ s), the
$\sigma^{f}/\sigma^{i}(t)$ curves show the well pronounced
maximum, which is practically absent on $P(t)$ curves (Fig.
\ref{f14}(a)). The interesting finding is that there is no direct
proportionality between the damage degree $P$ and relative
conductivity $\sigma^{f}/\sigma^{i}$. We can observe the obvious
decrease of $\sigma^{f}/\sigma^{i}$ even in the case when $P(t)$
practically does not change. This behaviour reflects the fact that
$\sigma^{f}/\sigma^{i}$ in the percolation phenomena depends not
only on damage degree $P$ but also on the spatial distribution of
the damaged cells over the system (Sahimi, 1994).

\section{Discussion}

The main hypothesis of the present work is that the electric field
damage of  a biological tissue is a phenomena of correlated
percolation. This phenomena is governed by the two key processes:
resealing of cells and moisture transfer inside of 
cellular structure. In the simulation model we try to use the
minimal number of parameters in order to imitate only the main
feature of this very complex phenomena, which comprises different
processes at various micro and macro-hierarchical levels of
membranes, cells, tissue structure, etc. We have found a
considerable difference between the damage kinetics at long-term
and short-term repetition times $\Delta t$ (see Fig. \ref{f9}(a)
for experimental data, and Fig. \ref{f11}(a) for simulation data).

In the case of long-term repetition times $\Delta t>\tau_{d}$ a
damage process in a system has a correlated character. The origin
of this behaviour is the following. For heterogenous tissue
structure in external electric field the largest 
gradient of field strengths arise near the already damaged  cells.
So the new cells (after the next PEF pulse) are destroyed mainly
near the previously damaged cells. In this case the damage
processes are spatially correlated and we really observe this
character of damage distribution at the simulated patterns (see
Fig. \ref{f12}(a)). In the case of short-term repetition times
$\Delta t<\tau_{d}$ a damage process in a system has a random
character. Cells are also destroyed after each PEF pulse, but 
their damage is latent and is "invisible" for the rest of the
system. 
 So, in this case the damage processes are spatially random (see
Fig. \ref{f12}(a)).

Unfortunately, we are presently unable to make more strict
comparison between the theoretical and experimental data as far as
we have no precise data for the sets of parameters used in the
present model. From the technological point of view, it is
preferable to use such mode of PEF treatment that allows to
achieve more homogeneous damage of a cellular material and the
maximal degree of damage. These conditions are subject for
optimization through variation of the PEF treatment mode, but we
also need here data on model parameters.

We should mention 
also some restrictions of the present model. This model  does not
take into account rather important details concerning the
structure of a cellular material, and particularly,  the large
scale spatial fluctuations of electrophysical properties inside
the sample. These fluctuations become experimentally visualized
after application of PEF with the long-term repetition times
$\Delta t$ (see Fig. \ref{f10}(a)) as the large scale percolative
channels. For this reason we do not discuss here the
experimentally observed effects of electric field strength $E$,
because this behaviour may be extremely sensitive to the
above-mentioned spatial fluctuations. Moreover, our model was
developed only for two-dimensional systems. It allows us to
simulate rather large-scale and realistic systems and avoid the
well-know finite size scaling effects (Watanabe, 1995). But
three-dimensional simulation
is, of course, more realistic.
All these restrictions should be overcome in future models
in order to attain more profound 
understanding of tissue damage kinetics.

\section{Conclusion}

The breakage of  a biological tissue under the PEF treatment may
be described as a correlated percolation phenomena, which is
controlled by two key processes: resealing of the cells and
moisture transfer inside of cellular structure. The breakage
kinetics is very sensitive to the repetition times $\Delta t$ of
PEF treatment. We observe correlated percolation patterns for the
case when $\Delta t$ exceeds the time of moisture transfer
processes $\tau_{d}$ and in the other case, when $\Delta
t<\tau_{d}$, the random percolation patterns are observed. The
long-term mode of pulse repetition times in PEF treatment allows
us to visualize the macroscopic percolation channels in the
sample.

\section*{Acknowledgements}

The authors would like to thank the ``Pole Regional Genie des
Procedes`` (Picardie, France) for providing financial support.
Authors also thank Dr. N. S. Pivovarova and Dr. A. B. Jemai for
help with the preparation of the manuscript.


\section*{References}


Abidor, I.G., Arakelyan, V.B., Chernomordik, L.V., V.B.,
Chizmadzhev Y.A., Pastushenko, V.F., \& Tarasevich, M.R. (1979).
Electric breakdown of bilayer membranes: 1. The main experimental
facts and their qualitative discussion. Bioelectrochemistry and
Bioenergetics, 6, 37-52.

Aguilera J. M., \& Stanley, D. W. (1999). Microstructural
principles of food processing and engineering. Aspen Publishers,
Gaithersburg.


Barbosa-C\'{a}novas, G.V., Pothakamury, U.R., Palou, E., \&
Swanson, B.G. (1998). Nonthermal Preservation of Foods (pp.
53-72). Marcel Dekker, New York.




Barsotti, L., \& Cheftel, J. C. (1998). Traitement des aliments
par champs electriques pulses. Science des Aliments, 18(6),
584-601.

Bazhal, M.I., \& Vorobiev, E.I. (2000). Electric treatment of
apple slices for intensifying juice pressing. Journal of the
Science of Food and Agriculture (in press).

Chang, D.C,  \& Reese, T. S, (1990). Changes in membrane structure
induced by electroporation as revealed by rapid-freezing electron
microscopy. Biophysical Journal, 58,1-12.

Chang, D. C., Chassy, B. M., Saunders, J. A., \& Sowers, A. E.
Eds. (1992). Guide to electroporation and electrofusion. Academic
Press, San Diego. 




Chizmadzhev, Y. A.,  Indenbom, A. V., Kuzmin, P. I., Galichenko,
S. V., Weaver, J. C.,  \& Potts, R. O. (1998). Electrical
properties of skin at moderate voltages: Contribution of
appendageal macropores. Biophysical Journal, 74(2), 843-856.

DeBruin K. A., \& Krassowska W. (1999). Modeling electroporation
in a single cell. I. Effects of field strength and rest potential.
Biophysical Journal, 77(3), 1213-1224.

Frank, D.J. , \& Lobb C.J.~ (1988). Highly efficient algorithm for
percolative transport studies in two dimensions. Physical Review,
E37, 302-307.

Gabriel, B.,  \& Teissie J. (1999) . Time courses of mammalian
cell electropermeabilization observed by millisecond imaging of
membrane property changes during the pulse. Biophysical Journal,
76(4), 2158-2165.

Gekkas, V. (1992). Transport phenomena of foods and biological
materials. Boca Raton, FL: CRC Press.

Grishko, A. A., Kozin, V. M., \& Chebanu, V. G. (1991).
Electroplasmolyzer for processing plant raw material. US Patent
no. 4723483.

Gulyi, I.S., Lebovka, N.I., Mank, V.V., Kupchik, M.P., Bazhal,
M.I., Matvienko, A.B., \& Papchenko, A.Y.~ (1994). Scientific and
practical principles of electrical treatment of food products and
materials. UkrINTEI, Kiev (in Russian).



Jemai, A.B. (1997). Contribution a l'etude de l'effet d'un
traitement electrique sur les cossettes de betterave a sucre.
Incidence sur le procede d'extraction. Th\`{e}se de Doctorat,
Universite de Technologie de Compiegne, Compiegne, France.

Kalmykova, I.S. (1993). Application of electroplasmolysis for
intensification of phenols extracting from the grapes in the
technologies of red table wines and natural juice. PhD Thesis,
Odessa Technological Institute of Food Industry, Odessa, Ukraine
(in Russian).


Knorr, D., \& Angersbach, A.(1998). Impact of high intensity
electric field pulses on plant membrane permeabilization. Trends
in Food Science \& Technology, 9, 185-191.

Knorr, D., Geulen, M., Grahl, T., \& Sitzmann, W. (1994). Food
application of high electric field pulses. Trends in Food Science
\& Technology, 5, 71-75.

Knorr, D., Heinz, V., Angersbach, A., \& Lee, D.-U. (2000).
Membrane permeabilization and inactivation mechanisms of
biological systems by emerging technologies. In Eighth
International Congress on Engineering and Food, Puebla, Mexico,
9-13 April, 2000, 15.

Kotnik, T, Miklavcic, D., \& Slivnik, T. (1998). Time course of
transmembrane voltage induced by time-varing electric fields -- a
method for theoretical analysis and its application.
Bioelectrochemistry and Bioenergetics, 45, 3-16.

Kulshrestha, S.A., \& Sastry, S.K. (1998). Electroporation of
vegetable tissue an ohmic heater. In IFT'S 1998 annual meeting,
Atlanta, 20-24 June 1998, 59C-2.


Lebedeva, N.E. (1987). Electric breakdown of bilayer lipid
membranes at short times of voltage effect. Biologicheskiye
Membrany, 4 (9), 994-998 (in Russian).

Lebovka, N.I., Bazhal, M.I., \& Vorobiev, E.I. (2000). Simulation
and experimental investigation of food material breakage using
pulsed electric field treatment. Journal of Food Engineering, 44,
213-223.

Lebovka, N.I., \& Mank, V.V. (1992). Phase diagram and kinetics of
inhomogeneous square lattice brittle fracture. Physica A, 181,
346-363.



McLellan, M.R., Kime, R.L., \& Lind, L.R. (1991).
Electroplasmolysis and other treatments to improve apple juice
yield. Journal of Science Food Agriculture, 57, 303-306.

Miyahara, K. (1985). Methods and apparatus for producing
electrically processed foodstuffs. US Patent no. 4522834.


Neumann, E., \& Boldt, E. (1990). Membrane electroporation: the
dye method to determine the cell membrane conductivity. In C.
Nikolau, D. Chapman, \& A.R. Liss (Eds.), Horisons in membrane
biotechnology (pp. 69-83). Wiley-Liss, New-York.

Neumann, E., Kakorin S., \& T\oe nsing, K. (1999). Fundamentals of
electroporative delivery of drugs and genes. Mini-review.
Bioelectrochemistry and Bioenergetics, 48, 3-16.

Neumann, E., T\oe nsing, K., Kakorin, S., Budde, P., \& Frey, J.
(1998). Mechanism of electroporative dye uptake by mouse B cells.
Biophysical Journal, 74, 98-108.



Papchenko, A.Y., Bologa, M.K., \& Berzoi (1988). Apparatus for
processing vegetable raw material. US Patent no. 4787303.

Press, W.H., Teukolsky, S.A., Vetterling, W.T., \& Flannery, B.P.
(1997). Numerical Recipes in Fortran 77: The Art of Scientific
Computing (vol.1). Cambridge University Press, Cambridge.

Rols, M.-P.,  \& Teissie, J. (1998). Electropermeabilization of
mammalian cells to macromolecules: Control by pulse duration.
Biophysical Journal, 75, 1415-1423.

Sahimi, M. (1994). Applications of Percolation Theory. Taylor and
Francis, London.

Saulis, G. (1997). Pore disappearance in a cell after
electroporation: theoretical simulation and comparison with
experiments. Biophysical Journal, 73, 1299-1309.


Scheglov, Ju.A., Koval, N.P., Fuser, L.A., Zargarian, S.Y.,
Srimbov, A.A., Belik, V.G., Zharik, B.N., Papchenko, A.Y.,
Ryabinsky, F.G., \& Sergeev, A.S. (1988). Electroplasmolyzer for
processing vegetable stock. US Patent no. 4753810.


Teissie, J., \&  Ramos, C. (1998).  Correlation between electric
field pulse induced long-lived permeabilization and fusogenicity
in cell membranes. Biophysical Journal, 74(4), 1889-1898.

Watanabe, M.S. (1995). Percolation with a periodic boundary
conditions: The effect of system size for crystallization in
molecular dynamics. Physical Review, E51(5), 3945-3951.

Weaver, J.C., \& Chizmadzhev, Yu.A. (1996). Theory of
electroporation: A review. Biolectrochemistry and Bioenergetics,
41(1), 135-160.

Weaver, J.C., Pliquett, U. \&  Vaughan T. (1999). Apparatus and
method for electroporation of tissue. US Patent no. 5983131.

Wouters, P. C., \& Smelt, J. P. P. M. (1997). Inactivation of
microorganisms with pulsed electric fields: Potential for food
preservation. Food Biotechnology, 11(3), 193-229.


Zimmermann, U. (1975). Electrical breakdown:
electropermeabilization and electrofusion. Reviews of Physiology
Biochemistry and Pharmacology, 105, 176-256.





\begin{figure}[tbp]
\caption{Typical high
resolution scanning electron micrograph of the apple sample. The
"WET" chamber mode was employed that allowed observation of
hydrated apple specimens in their natural state.} \label{f0}
\end{figure}

\begin{figure}[tbp]
\caption{Schematic
representation of the experimental set-up used in the study of
pulsed electric field treatment of apple slices.} \label{f1}
\end{figure}

\hspace*{2cm}

\begin{figure}[tbp]
 \caption{The example
of the initial apple slab conductivity changes caused by the
freshly cut boundary regions. The dashed line shows the transition
time which always is skipped before the PEF treatment.} \label{f2}
\end{figure}

\begin{figure}[tbp]
 \caption{ Survival
probability for a cell $S$ versus normalised transmembrane voltage
$u^{\ast }=u/u_{o}$ calculated using Eq. (\ref{e03}) at $\Delta
u^{\ast}=0.26$, $\tau _{\infty }=3.7\times10^{-7}$ s, and $dt=100$
$\mu $s. Here, the dotted straight line is a tangent line to a
curve $S(u^{\ast })$ in the midpoint $u^{\ast }=1$ (is shown by
the small circle), and the dashed line corresponds to the
normalised density distribution function
$S^{\prime}/S^{\prime}{_{max}}$, where $S^{\prime}=dS/du^{\ast}$.}
\label{f3}
\end{figure}

\begin{figure}[tbp]
 \caption{The
two-dimensional model of the cellular material structure. Each
cell is represented by a node with four conducting bonds. Here
$d_{c}$ is a mean radius of a cell, $d_{m}$ is a membrane
thickness, $L=2d_{c}N$ is the total thickness of a sample, and $U$
is an external voltage applied to the sample.} \label{f4}
\end{figure}

\begin{figure}[tbp]
\caption{ This
explains the procedure of the resistors interchange when the cells
are damaged in vertical (a) or horizontal (b) directions. Here
resistances $r_{i}$ and $r_{d}$ correspond to the intact and
damaged cells and are defined by the Eq. (\ref{e08}) and Eq.
(\ref{e14}), respectively.} \label{f5}
\end{figure}

\begin{figure}[tbp]
\caption{ The
calculation of the chosen site potential $u_{x,y}^{n}$ according
to successive relaxation scheme. For this particular case the
central intact cell is surrounded by three intact and one damaged
cells. The bond conductivities are given by
$G_{1}=(r(x,y)+r(x,y+1))^{-1}=(2r^{i})^{-1}$,
$G_{2}=(r(x,y)+r(x-1,y))^{-1}=(2r^{i})^{-1}$,
$G_{3}=(r(x,y)+r(x,y-1))^{-1}=(2r^{i})^{-1}$, and
$G_{4}=(r(x,y)+r(x+1,y))^{-1}=(r^{i}+r^{d})^{-1}$, respectively.}
\label{f6}
\end{figure}

\begin{figure}[tbp]
\caption{ The example
of the simulated breakdown kinetics: degree of breakdown $P$ and
effective conductivity $\sigma$ of the system versus number of
pulses $n$. Here $t_{i}=1$ ms is a pulse duration, $\Delta
t=5t_{i}=5\times10^{-3}$ is a pulse repetition time,
$\tau_{r}=10^{-2}$ s  is a resealing time, and $\tau_{d}$ is a
mass transfer process time.  The calculations are performed at
$E^{\ast }=0.75$, $m=1$ (a case of complete resealing) and
$c=0.1$. Here, all the parameters are chosen only with the purpose
of clear illustration of the work of algorithm. } \label{f7}
\end{figure}

\begin{figure}[tbp]
 \caption{Examples of relative conductivity
$\protect\sigma^{f}/\protect\sigma ^{i}$ of apple slabs versus
time $t$ dependencies at different values of the electric field
strength $E=500$ V cm$^{-1}$ and $E=200$ V cm$^{-1}$ and pulse
protocols: $t_i=1$ ms, $n=1-15$, $\Delta t=60$ s. } \label{f8}
\end{figure}

\begin{figure}[tbp]
 \caption{
Relative conductivity $\protect\sigma^{f}/\protect\sigma ^{i}$ of
apple slabs versus number of pulses $n$ at different values of the
electric field strength $E=200$ V cm$^{-1}$ and $E=500$ V
cm$^{-1}$ and two pulse protocols: $t_i=1$ ms, $n=1-15$, $\Delta
t=60$ s (protocol I), and $\Delta t=10$ ms (protocol II). In all
the cases the value of $\protect\sigma^{f}/\protect\sigma ^{i}$
was measured at time $t_{m}=10 s$ after the end of PEF treatment.
All the experiments were repeated five times. The error bars
represent standard data deviations.} \label{f9}
\end{figure}

\begin{figure}[tbp]
\caption{
Photographs which illustrate the structure changes of the apple
slabs after PEF treatment at $E=500$ V cm$^{-1}$, $t_i=1$ ms,
$n=10$ and different pulse repetition time $\Delta t=60$ s (a,
protocol I), and $\Delta t=10$ ms (b, protocol II). }

\label{f10}
\end{figure}

\begin{figure}[tbp]
 \caption{
Calculated curves of relative conductivity
$\protect\sigma^{f}/\protect\sigma^{i}$ versus time $t$ (a)  and
 number of pulses $n$ (b) for two different pulse repetition
times $\Delta t=60$ s (protocol I) and  $\Delta t=10$ ms (protocol
II). The following values parameters were used in these
calculations: $t_{i}=1$ ms, $m=0.1$, $c=0.1$, $\tau_{r}=10$ s,
$\tau_{d}=0.5$ s, $n=1-15$ and $E^{*}=0.65$.} \label{f11}
\end{figure}

\begin{figure}





\caption{Simulated breakage patterns for two different pulse
repetition times $\Delta t=60$ s (a, protocol I) and $\Delta t=10$
ms (b, protocol II). We have used the following values of
parameters in these calculations: $t_i=1$, $m=0.1$, $c=0.1$,
$\tau_{r}=10$ s, $\tau_{d}=0.5$ s and $E^{*}=0.65$. In each
pattern we display only those cells, which were broken after the
$n$-th pulse. } \label{f12}
\end{figure}

\begin{figure}[tbp]
\caption{Calculated
$(\sigma^{f}/\sigma^{i})_{t_{m}}\rightarrow\infty $
 versus $\Delta t$ dependencies at different values of
 $m$ (a, $C=0.1$ and $\tau_{r}=10$ s)
and $c$ (b, $\tau_{r}=\infty$, i.e., when the resealing is
absent). All calculation were performed at $\tau_{d}=0.5$ s. The
relevant repetition time when $\Delta t\sim \tau_{r}$ is shown by
dashed line.} \label{f13}
\end{figure}

\begin{figure}[tbp]
 \caption{The simulated kinetics of damage degree $P$
and relative conductivity $\sigma^{f}/\sigma^{i}$ for different
values of $\tau_{d}$, $\tau_{r}=10$ s (a) and $\tau_{r}=\infty$
(b). Arrows show the direction of $\tau_{d}$ increase. All the
calculations were performed for the following values of parameters
$m=0.05$, $c=0.1$ s and $E^{*}=0.65$.} \label{f14}
\end{figure}


\end{document}